\documentstyle[11pt]{article}
\begin{document}
\vskip 22pt

\vspace{0.5cm}

\centerline{\Large \bf  BOSE SYMMETRY INTERFERENCE EFFECTS OF $4\pi$ FINAL
STATES }
\vspace{8pt}
\centerline{\Large \bf  }
\vspace{1cm}

\centerline{ Jie Chen$^{a,b,c}$,
             Xue-Qian Li$^{a,b,c}$, Bing-Song Zou$^{a,c}$}
\vspace{0.5cm}
{\small
\vspace{8pt}
\flushleft{  a. CCAST(World Laboratory), P.O.Box 8730, Beijing 100080, China}
\vspace{8pt}
\flushleft{  b. Department of Physics, Nankai University, Tianjin
300071, China}
\vspace{8pt}
\flushleft{  c. Institute of High Energy Physics, Academia Sinica,
 P.O.Box 918(4), Beijing 100039, China}
}

\date{\today}

\vspace{0.6cm}

\begin{center}
\begin{minipage}{11cm}

\noindent{\bf Abstract}

\vspace{8pt}

{\small

\noindent
We carefully analyze the relative branching ratios of $4\pi$ final states
$\pi^+\pi^-\pi^+\pi^-$, $\pi^+\pi^-\pi^0\pi^0$ and $\pi^0\pi^0\pi^0\pi^0$,
from various resonances of $J^{PC}=0^{++}$, $0^{-+}$, $2^{++}$. We find
that the Bose symmetry interference effects would make their ratios to
obviously differ from the naive counting values without considering these
effects. The results should be applied to estimate correctly various
$4\pi$ decay branching ratios of relevant  resonances.

\vspace{1cm}

{\bf PACS number(s): 14.40Cs, 13.25Jx, 13.75Cs, 13.25Gv}
}

\end{minipage}
\end{center}
\baselineskip 22pt

\vspace{1cm}

\noindent {\bf I. Introduction}

\vspace{0.5cm}

In the energy range of $1\sim 2.5$ GeV, there exist very rich hadronic
resonance spectra, including possible $0^{++}$, $0^{-+}$,
and $2^{++}$ glueballs. An important source of information about the
nature of these resonances is their various decay branching ratios.
Among the observed decay modes for the $0^{++}$, $0^{-+}$ and
$2^{++}$ resonances, the $4\pi$ final state is a very important
one\cite{PDG}.

There are three kinds of $4\pi$ final states:
$\pi^+\pi^-\pi^+\pi^-$, $\pi^+\pi^-\pi^0\pi^0$ and $\pi^0\pi^0\pi^0\pi^0$.
Usually, due to specialty and limitation of each detector and other
reasons, one experiment is only good at studying one kind of the $4\pi$ final
states. For example, the Crystal Barrel (CBAR) detector is particularly
good at studying neutral final states and therefore studied resonances
decaying into $\pi^0\pi^0\pi^0\pi^0$ final state\cite{Scott};
the BES (Beijing Spectrometer) detector is good at detecting charged
particles and only has studied $J/\Psi$ radiative decaying into
$\pi^+\pi^-\pi^+\pi^-$ final state\cite{Bai}. The problem is that from the
measured rate for one kind of the $4\pi$ final states how to deduce the rates
for other two kinds of the $4\pi$ final states.

The $4\pi$ final states are usually produced via 2-meson intermediate
states $M_1$ and $M_2$, namely the parent-resonance decays into two mesons
which then result in the $4\pi$ final states:
$$M\rightarrow M_1+M_2\rightarrow 4\pi.$$
The parent-meson which we are interested in are $0^{++},0^{-+}$ and
$2^{++}$ etc., because they have the same quantum numbers as glueballs
which are under intensive discussions at present. $M_1$ and $M_2$ can be
various mesons among which $\sigma\sigma,\rho\rho$ and $f_2\sigma$
are the most possible candidates\cite{Scott,Bai} and therefore discussed
in this work.

Naive counting for the $4\pi$ final states from simple isospin
decomposition would result in
\begin{eqnarray}
&& \Gamma(M\rightarrow\pi^+\pi^-\pi^0\pi^0):\Gamma(M\rightarrow\pi^+\pi^-
\pi^+\pi^-): \Gamma(M\rightarrow\pi^0\pi^0\pi^0\pi^0) \nonumber\\
&&=4:4:1,
\;\;\;  {\rm for \;f_2\sigma,\sigma\sigma \; intermediate\; states};\\
&& \Gamma(M\rightarrow\pi^+\pi^-\pi^0\pi^0):\Gamma(M\rightarrow\pi^+\pi^-
\pi^+\pi^-): \Gamma(M\rightarrow\pi^0\pi^0\pi^0\pi^0) \nonumber\\
&&=2:1:0,  \;\;\;  {\rm for \;\rho\rho \;intermediate\; states},
\end{eqnarray}
where all interference effects are neglected.

Since all pions can be treated as identical particles and each mode has the
same production amplitude up to an SU(2) factor, so unless all the
interference terms cancel each other after integration over the invariant
phase space of final states, their effects can be important. In earlier
literatures, the naive counting was employed to evaluate production rate
of one mode from others. Even though this counting way is simple and valid
in certain cases, it may bring up remarkable errors in some cases.

In this work, we carefully analyze the relative ratios of
$B(M\rightarrow\pi^+\pi^-\pi^0\pi^0):B(M\rightarrow\pi^+\pi^-
\pi^+\pi^-): B(M\rightarrow\pi^0\pi^0\pi^0\pi^0)$ by including
interference terms precisely for various masses of the parent meson of
$0^{++},0^{-+},2^{++}$, and intermediate 2-meson states.

In Sec.II, we present the formulation for analysis; the numerical results
and discussion are given in Sec.III. \\

\noindent{\bf II. Formulation}

\vspace{0.5cm}

For $M\rightarrow \pi_1(p_1)\pi_2(p_2)\pi_3(p_3)\pi_4(p_4)$ where $p_i's\;
(i=1...4)$ are the four-momenta of the four produced pions and we
use notation
$$p_{ab}\equiv p_a+p_b,\;\;\;\;a,b=1,...,4,\;{\rm and}\; a\neq b.$$
The propagators take the Breit-Wigner form \cite{Lan,Chung}
\begin{eqnarray}
\label{prop}
F_{ab} &=& {-i\over p_{ab}^2-m^2+i\Gamma m} \;\;\;
{\rm for \;scalar\; mesons}; \nonumber\\
D_{ab}^{\alpha\beta} &=& {i\over p_{ab}^2-m^2+i\Gamma m}\tilde
g^{\alpha\beta}\;\;\;  {\rm for\; massive\; vector}; \nonumber \\
D^{\alpha\beta\gamma\delta}_{ab} &=& {-i\over p_{ab}^2-m^2+i\Gamma m}
[{1\over 2}(\tilde g^{\alpha\gamma}\tilde g^{\beta\delta}+
\tilde g^{\alpha\delta}\tilde g^{\beta\gamma})-{1\over 3}
\tilde g^{\alpha\beta}\tilde g^{\gamma\delta}] \nonumber \\
&& {\rm for\; spin-2\; tensor\; meson},
\end{eqnarray}
where
\begin{equation}
\tilde g^{\alpha\beta}\equiv -g^{\alpha\beta}+{p_{ab}^{\alpha}p_{ab}^{\beta}
\over m^2},
\end{equation}
and $m$ is the mass of the concerned intermediate meson of spin 0 or 1 or 2.
In the following we explicitly present the expressions with $M\rightarrow
M_1+M_2\rightarrow 4\pi$ for various $M,M_1,M_2$ identities.\\

1. Decay of $M (0^{++})\rightarrow 4\pi$.

To investigate the interference effects of $M\rightarrow 4\pi$, we
distinguish the processes caused by different intermediate states.
Here we first ignore possible interferences between $M\to\sigma\sigma
\rightarrow 4\pi$ and $M\rightarrow\rho\rho\rightarrow 4\pi$ and
later we will show that except for special cases, it is legitimate.  Then
we argue that the conclusion can be generalized to most situations.

(a) The squares of amplitudes corresponding to $\sigma\sigma$ intermediate
state are
\begin{eqnarray}
\label{sig}
|M|^2 &=& {g^2\over 2}|F_{12}^{\sigma}F_{34}^{\sigma}|^2 \;\;\;\;
 {\rm for}\;f_0\rightarrow
\pi^+\pi^-\pi^0\pi^0;
\nonumber\\
|M|^2 &=& {g^2\over 4}|F_{12}^{\sigma}F_{34}^{\sigma}+
F_{14}^{\sigma}F_{32}^{\sigma}|^2 \;\;\;\;
 {\rm for}\; f_0\rightarrow
\pi^+\pi^-\pi^+\pi^- ;
\nonumber\\
|M|^2 &=& {g^2\over 24}|F_{12}^{\sigma}F_{34}^{\sigma}+
F_{13}^{\sigma}F_{24}^{\sigma}+
F_{14}^{\sigma}F_{32}^{\sigma}|^2 \;\;\;\;
{\rm for}\; f_0\rightarrow \pi^0\pi^0\pi^0\pi^0.
\end{eqnarray}

(b) via $\rho\rho$ intermediate states.
\begin{eqnarray}
\label{rh}
|M|^2 &=& {g^{'2}\over 2}|(p_1-p_2)\cdot (p_3-p_4)F_{12}^{\rho}F_{34}^{\rho}
+ (p_1-p_4)\cdot (p_3-p_2)F_{14}^{\rho}F_{32}^{\rho}|^2\nonumber \\
 &&{\rm for}\;f_0\rightarrow \pi^+\pi^0\pi^-\pi^0; \nonumber \\
|M|^2 &=& {g^{'2}\over 4}|(p_1-p_2)\cdot (p_3-p_4)F_{12}^{\rho}F_{34}^{\rho}
+ (p_1-p_4)\cdot (p_3-p_2)F_{14}^{\rho}F_{32}^{\rho}|^2 \nonumber \\
 &&{\rm for}\;f_0\rightarrow \pi^+\pi^-\pi^+\pi^-,
\end{eqnarray}
where $p_i's$ are the momenta of the outgoing pions and
$F_{ij}^{\sigma},F_{ij}^{\rho}$ are the propagators of $\sigma-$meson and
$\rho-$meson respectively. It is noted that there is no process
$f_0\rightarrow\rho\rho\rightarrow 4\pi^0$,
because $\rho^0\rightarrow \pi^0\pi^0$ is forbidden by isospin symmetry.\\

2. Decay of $0^{-+}$ mesons.

It is obvious that $0^{-+}\rightarrow \sigma\sigma$ is forbidden
by parity and angular-momentum conservations, it decays into
$4\pi$ only via $\rho\rho$ intermediate states.
\begin{eqnarray}
\label{0np}
|M|^2&=&{g^{'2}\over 2}|\epsilon_{\mu\nu\lambda\rho}
(p_1^{\mu}p_2^{\nu}p_3^{\lambda}p_4^{\rho}F_{12}^{\rho}
F_{34}^{\rho}-p_1^{\mu}p_4^{\nu}p_3^{\lambda}p_2^{\rho}
F_{14}^{\rho}F_{32}^{\rho})|^2 \nonumber\\
&& {\rm for}\;0^{-+}\rightarrow \pi^+\pi^0\pi^-\pi^0;\nonumber\\
|M|^2&=&{g^{'2}\over 4}|\epsilon_{\mu\nu\lambda\rho}
(p_1^{\mu}p_2^{\nu}p_3^{\lambda}p_4^{\rho}F_{12}^{\rho}
F_{34}^{\rho}-p_1^{\mu}p_4^{\nu}p_3^{\lambda}p_2^{\rho}
F_{14}^{\rho}F_{32}^{\rho})|^2 \nonumber\\
&& {\rm for}\;0^{-+}\rightarrow \pi^+\pi^-\pi^+\pi^-.
\end{eqnarray}

3. Decay of $2^{++}$ mesons.

(a) Via $\sigma\sigma$ intermediate states,
\begin{eqnarray}
|M|^2 &=& {g^{''2}\over 2}|\sqrt{{1\over 6}}(-r^2+3r_Z^2)F_{12}^{\sigma}
 F_{34}^{\sigma}|^2\nonumber\\
&& { \rm for}\; 2^{++}\rightarrow \pi^+\pi^-\pi^0\pi^0; \nonumber \\
|M|^2 &=& {g^{''2}\over 4}|\sqrt{{1\over 6}}(-r^2+3r_Z^2)
F_{12}^{\sigma}F_{34}^{\sigma} +\sqrt{{1\over 6}}(-r^{''2}+3r_Z^{''2})
F_{14}^{\sigma}F_{32}^{\sigma}|^2 \nonumber\\ && {\rm for}\;
2^{++}\rightarrow \pi^+\pi^-\pi^+\pi^-; \nonumber \\
|M|^2 &=& {1\over 24}g^{''2}|\sqrt{{1\over
6}}(-r^2+3r_Z^2)F_{12}^{\sigma}F_{34}^{\sigma}+
\sqrt{{1\over 6}}(-r^{'2}+3r_Z^{'2})F_{13}^{\sigma}F_{24}^{\sigma}+
\nonumber \\ && \sqrt{{1\over 6}}
(-r^{''2}+3r_Z^{''2})F_{14}^{\sigma}F_{32}^{\sigma}|^2\;\;\;\;
 {\rm for}\; 2^{++}\rightarrow \pi^0\pi^0\pi^0\pi^0,
\end{eqnarray}
where
$$r\equiv p_{12}-p_{34}, \;\;\; r^{'}\equiv p_{13}-p_{24}, \;\;\;
r^{''}\equiv p_{14}-p_{32}.$$

(b) Via $\rho\rho$ intermediate states.

For $2^{++}$ decays, $\sigma\sigma$ production can only occur at d-wave,
which is more suppressed, therefore, the $\rho\rho$ mode may dominate in
the $2^{++}\rightarrow 4\pi$ decays. The amplitudes are
\begin{eqnarray}
|M|^2 &=&
{g^{''2}\over 2}|(-p_{12}^xp_{34}^x-p_{12}^yp_{34}^y+2p_{12}^zp_{34}^z)
F_{12}^{\rho} F_{34}^{\rho}+(-p_{14}^xp_{32}^x-p_{14}^yp_{32}^y
+2p_{14}^zp_{32}^z)F_{14}^{\rho}F_{32}^{\rho}|^2 \nonumber\\
&&  {\rm for}\; 2^{++}\rightarrow\pi^+\pi^0
\pi^-\pi^0;\nonumber\\
|M|^2 &=&
{g^{''2}\over 4}|(-p_{12}^xp_{34}^x-p_{12}^yp_{34}^y+2p_{12}^zp_{34}^z)
F_{12}^{\rho} F_{34}^{\rho}+(-p_{14}^xp_{32}^x-p_{14}^yp_{32}^y
+2p_{14}^zp_{32}^z)F_{14}^{\rho}F_{32}^{\rho}|^2 \nonumber\\
&& {\rm for}\; 2^{++}\rightarrow\pi^+\pi^-\pi^+\pi^-.
\end{eqnarray}

4. Decay of $2^{++}\rightarrow 4\pi$ via $f_2\sigma$ intermediate states.

Here we have a simplified expression for the decay modes
instead of using the propagator given in eqs.(\ref{prop}) as
\begin{equation}
|M|^2=|(-T^{11}-T^{22}+2T^{33})_{(ab)}F_{f_2}(ab)F_{\sigma}(cd)|^2,
\end{equation}
where
\begin{equation}
T^{ii}=q^iq^i+{1\over 3}(1+p^ip^i/M_{f_2}^2)|\vec q|^2,
\end{equation}
and
$$ p_{ab}=p_a+p_b,\;\; q_{ab}=p_a-p_b,\;\; F_{f_2}={1\over M_{f_2}^2-s_{ab}
-iM_{f_2}\Gamma_{f_2}},\;\; s_{ab}=p_{ab}^2.$$
The subscript $(ab)$ denotes the argument indices in the tensor
$T^{ii}$, Thus
\begin{eqnarray}
|M|^2 &=& {1\over 2}|(-T^{11}-T^{22}+2T^{33})_{12}
F_{f_2}(12)F_{\sigma}(34)|^2\nonumber \\
&& {\rm for}\;2^{++}\rightarrow f_2\sigma\rightarrow \pi^+\pi^-\pi^0\pi^0;
\nonumber \\
|M|^2 &=& {1\over 4}|(-T^{11}-T^{22}+2T^{33})_{12}F_{f_2}(12)F_{\sigma}(34)
+(-T^{11}-T^{22}+2T^{33})_{14}F_{f_2}(14)F_{\sigma}(32)|^2
\nonumber \\
&& {\rm for}\;2^{++}\rightarrow f_2\sigma\rightarrow \pi^+\pi^-\pi^+\pi^-;
\nonumber\\
|M|^2 &=& {1\over
24}|(-T^{11}-T^{22}+2T^{33})_{12}F_{f_2}(12)F_{\sigma}(34)+
(-T^{11}-T^{22}+2T^{33})_{13}F_{f_2}(13)F_{\sigma}(24)+\nonumber\\
&& (-T^{11}-T^{22}+2T^{33})_{14}F_{f_2}(14)F_{\sigma}(32)|^2 \nonumber\\
&& {\rm for}\;2^{++}\rightarrow f_2\sigma\rightarrow \pi^0\pi^0\pi^0\pi^0.
\end{eqnarray}
\\

5. The interference between channels with
$\sigma\sigma$ and $\rho\rho$ intermediate states
in $M\rightarrow 4\pi$.

(a) Above, we have ignored possible interference between channels  with
$\sigma\sigma$ and $\rho\rho$ intermediate states for the $4\pi$ final
states, just because we assume that one of the two modes would overwhelm
over the other. This allegation might deviate from reality. So in this
subsection, we study this interference effects. As an example, we only
concentrate on the $0^{++}$
decays. Then we have the squares of amplitudes as
\begin{eqnarray}
\label{sr}
|M|^2 &=& \frac{1}{2}|g^{'}(p_1-p_2)\cdot (p_3-p_4)F_{12}^{\rho}F_{34}^{\rho}
+g^{'}(p_1-p_4)\cdot (p_3-p_2)F_{14}^{\rho}F_{32}^{\rho}
+gF_{13}^{\sigma}F_{24}^{\sigma}|^2 \nonumber \\
&& {\rm for}\; f_0\rightarrow\pi^+\pi^0\pi^-\pi^0;\nonumber\\
|M|^2 &=& {1\over 4}|g^{'}(p_1-p_2)\cdot (p_3-p_4)F_{12}^{\rho}F_{34}^{\rho}
+g^{'}(p_1-p_4)\cdot (p_3-p_2)F_{14}^{\rho}F_{32}^{\rho}+
g(F_{12}^{\sigma}F_{34}^{\sigma}
+F_{14}^{\sigma}F_{32}^{\sigma})|^2 \nonumber \\
& &{\rm for}\; f_0\rightarrow
\pi^+\pi^-\pi^+\pi^-;\nonumber\\
|M|^2 &=& {1\over 24}|g F_{12}^{\sigma}F_{34}^{\sigma}+
g F_{13}^{\sigma}F_{24}^{\sigma}+
gF_{14}^{\sigma}F_{32}^{\sigma}|^2
\nonumber \\
&&{\rm for} \; f_0\rightarrow \pi^0\pi^0\pi^0\pi^0.
\end{eqnarray}

(b) As an illustration let us study the decay of $f_0(1750)$, because there
are data available for $f_0(1750)\rightarrow \rho\rho$ and $\sigma\sigma$
\cite{Bugg}. The partial decay widths are
\begin{eqnarray}
\label{gam}
\Gamma(f_0\rightarrow\sigma\sigma) &=& {g_f^2\over 16\pi M_f}
(1-{4m_{\sigma}^2\over M_f^2})^{1/2},\nonumber \\
\Gamma(f_0\rightarrow\rho\rho) &=& {g_f^{'2}\over 16\pi M_f}
(1-{4m_{\rho}^2\over
M_f^2})^{1/2}[3+{1\over 4m_{\rho}^4}(M_f^4-4M_f^2m_{\rho}^2)],\nonumber \\
\Gamma(\rho\rightarrow 2\pi) &=& {g_{\rho}^2m_{\rho}\over 48\pi}
(1-{4m_{\pi}^2\over m_{\rho}^2})^{3/2},\nonumber \\
\Gamma(\sigma\rightarrow\pi\pi) &=& {g_{\sigma}^2\over 16\pi m_{\sigma}}
(1-{4m_{\pi}^2\over m_{\sigma}^2})^{1/2}.
\end{eqnarray}
So we have
\begin{equation}
\label{coup}
g\equiv g_f\cdot g_{\sigma}^2,\;\;\;\;\; g^{'}\equiv g^{'}_f
\cdot g_{\rho}^2.
\end{equation}
By setting $\Gamma^{th}=\Gamma^{exp}$, we have obtained all the coupling
constants straightforwardly.\\

\noindent{\bf III. Numerical results and discussion}

\vspace{0.5cm}

We have employed the Monte-Carlo program to carry out the
calculations of the widths. And in the practical calculation, we need to
multiply the propagators of vector and tensor $F_\rho$ and $F_{f_2}$ in
all the equations by the Blatt-Weisskopt barrier factor
$B_l(p)$\cite{Chung,von} ($B_0(p)=1$), which is widely used in
partial-wave analyses.
Namely, in our program the $F_\rho $ is replaced by
 $F_\rho B_l(p)$ and $ F_{f_2}$ by $F_{f_2} B_2(p)$
respectively. For the $\sigma-$propagator, there are various forms.
Here we only use two typical ones. The first is \cite{D.V.Bugg},
\begin{eqnarray}
\label{fsgm}
F_\sigma&=&\frac{1}{M_\sigma^2-s-iM_\sigma(\Gamma_1(s)+\Gamma_2(s))},
\\ {\rm where}\nonumber\\
&&\Gamma_1(s)=G_1
\frac{\sqrt{1-4m_\pi^2/s}}{\sqrt{1-4m_\pi^2/M_\sigma^2}}
\frac{s-m_\pi^2/2}{M_\sigma^2-m_\pi^2/2}
e^{-(s-M_\sigma^2)/4\beta^2},\nonumber\\
&&\Gamma_2(s)=G_2{\sqrt{1-16m_\pi^2/s}\over 1+exp(\Lambda(s_0-s))}\cdot
{1+exp(\Lambda(s_0-M^2_\sigma))\over \sqrt{1-16m_\pi^2/M^2_\sigma}}
\end{eqnarray}
with $M_\sigma=1.067$ GeV, $G_1=1.378$ GeV, $\beta=0.7$ GeV,
$G_2=0.0036$ GeV, $\Lambda=3.5$ GeV$^{-2}$ and $s_0=2.8$ GeV$^2$.

The second is \cite{zou},
\begin{eqnarray}
\label{fsgm2}
F_\sigma&=&\frac{e^{2i\phi}-1}{2i}+\frac{g_1\rho_1e^{2i\phi}}
{M_R^2-s-i(\rho_1g_1+\rho_2g_2)},\nonumber\\
\\ {\rm with}\nonumber\\
&& e^{2i\phi}=\frac{1+a_1s+a_2s^2+i\rho_1[b_1(s-M_\pi^2/2)+b_2s^2]}
{1+a_1s+a_2s^2-i\rho_1[b_1(s-M_\pi^2/2)+b_2s^2]},
\end{eqnarray}
where $ a_1=-0.3853 GeV^{-2}$, $a_2=-0.4237 GeV^{-4}$, $b_1=-3.696
GeV^{-2}$, $b_2=-1.462 GeV^{-4}$, $g_1=0.1108$,
$g_2=0.4229$, $M_R=0.9535 GeV$, $\rho_1=\sqrt{1-4m_\pi^2/s}$,
$\rho_2=\sqrt{1-4m_\kappa^2/s}$, and s is the invariant mass squared of
the system. This one is in fact the full $\pi\pi$ S-wave
scattering amplitude corresponding to CERN-M\"unich $\pi\pi$ S-wave phase
shifts\cite{CM} and is very close to the AMP amplitude{\cite{AMP}, and
hence includes contributions from several $0^{++}$ resonances.

It is noted that when we only concern the relative values of the branching
ratios, as in most parts of this work we do not need the concrete
coupling constants in eqs.(\ref{0np}) through (\ref{gam}) and our
numerical results of
branching ratios may differ from the real
values by a constant. Obviously this does not affect our conclusion at all.

Below  we will present our results in graphs and make also discussions. In
next section, we will summarize what we have learned from this investigation.

1. In Fig.1 we present the relative branching ratios of
$B(0^{++}\rightarrow \sigma\sigma\rightarrow 4\pi)$ which are
normalized  by $B(0^{++}\rightarrow \sigma\sigma\rightarrow
4\pi^0)$. It is obvious that the ratios of
$B(\pi^+\pi^-\pi^0\pi^0):B(\pi^+\pi^-\pi^+\pi^-):B(\pi^0\pi^0\pi^0\pi^0)$
decline from the naive counting 4:4:1. The
curves $\pi^+\pi^-\pi^0\pi^0 a$ and $\pi^0\pi^0\pi^0\pi^0 b$ are evaluated
in terms of eq.(\ref{fsgm}) for the $\sigma-$propagator while the others
correspond to eq.(\ref{fsgm2}).
The same conventions  apply to Fig.4 and Fig.6.

2. To study the significance of the interference effects in $M\to 4\pi$,
we define a quantity $R$ as
\begin{equation}
R=\frac{\int[LIPS]\sum_i|A_i|^2}{\int[LIPS]|\sum_iA_i|^2},
\end{equation}
where the sum runs over all channels which contribute to the same $4\pi$
products, so the channels may interfere among each other and the
integration is carried out over the Lorentz Invariant Phase Space (LIPS).

Fig.2 gives the R-values for $0^{++}\to\rho\rho\to\pi^+\pi^-\pi^0\pi^0$
and $0^{++}\rightarrow \rho\rho\rightarrow\pi^+\pi^-\pi^+\pi^-$,
because $0^{++}\rightarrow \rho\rho\rightarrow\pi^0\pi^0\pi^0\pi^0$ is
forbidden.

3. Fig.3 is for $0^{-+}\rightarrow\rho\rho\rightarrow 4\pi$ in analog to
Fig.2.

4.  Fig.4 is for $2^{++}\rightarrow \sigma\sigma\rightarrow 4\pi$
and Fig.5 is for $2^{++}\rightarrow \rho\rho\rightarrow 4\pi$.

5. Recently, the BES collaboration discovered a new possible
channel $f_2\sigma$ in the $2^{++}-$resonance decay, thus as an
intermediate state, it can also contribute to the $4\pi$ final
states. Fig.6 shows the relative branching ratios of $2^{++}\rightarrow
f_2\sigma\rightarrow \pi^+\pi^-\pi^0\pi^0,\pi^+\pi^-\pi^+\pi^-,
\pi^0\pi^0\pi^0\pi^0$ respectively where as usual they are also
normalized by $B(2^{++}\rightarrow \pi^0\pi^0\pi^0\pi^0)$.

6. As aforementioned, we deliberately ignore the interference
among different intermediate channels. It is true if one of the
channels prevails over the others. Here we study the interference
between $\sigma\sigma$ and $\rho\rho$ intermediate channels in
$0^{++}\rightarrow 4\pi$ decays. This theoretical estimation
depends on the effective couplings $g^{'}$ and $g$ which are
formulated in eq.(\ref{coup}). In Fig.7 and Fig.8 we deal with the
$\pi^+\pi^-\pi^0\pi^0$ and $\pi^0\pi^0\pi^0\pi^0$ final states
respectively. And for the convenience, we compute
$|M|_{\sigma\sigma}^2  {\rm\; according  \;to \;eq.(\ref{sig})}$ and
$|M|_{\rho\rho}^2 {\rm \;according \;to \; eq.}(\ref{rh})$, then obtain the
interference term as
$B_I=$$(|M|^2-|M|_{\rho\rho}^2-|M|_{\sigma\sigma}^2)$ for
various masses of the parent meson where the formula for $|M|^2$
is given in eq.$(\ref{sr})$.
 In Fig.7,
g'/g takes 4.5 whereas 8.5 in Fig.8,  we choose
the form(\ref{fsgm}) for a  $\sigma-$propagator.

For the $f_0(1500)$, $f_0(1750)$ and $f_0(2100)$, the $\sigma\sigma$
intermediate state dominates over the $\rho\rho$ intermediate
states\cite{Bai,Bugg}. Hence the interference between $\sigma\sigma$ and
$\rho\rho$ intermediate states is negligible for these states.

Now we apply our results to estimate the branching ratios of the
channels which are not measured yet.

Below we tabulate the numerical results for some branching ratios of
$J/\Psi$ radiative decays to $4\pi$ states where only one
of the three $4\pi$-modes ($\pi^+\pi^-\pi^+\pi^-$)
is experimentally measured. In  table 1,  the first column
contains the values of  $B(\pi^+\pi^-\pi^+\pi^-)$ measured by the BES
collaboration \cite{Bai}, while the other two
columns are for the ones evaluated in terms of our scheme where
interference effects are carefully considered.
The table 2 is similar but based on the MARKIII data
\cite{Bugg}.

\begin{center}

Table 1. Branching ratios of $J/\Psi$ radiative decays to $4\pi$ based on
the BES data

\vspace{0.3cm}

\begin{tabular}{|l|l|l|l|l|}\hline
parent meson &  $B(\pi^+\pi^-\pi^+\pi^-)$ (measured)
& $B(\pi^+\pi^-\pi^0\pi^0)$ & $B(\pi^0\pi^0\pi^0\pi^0)$&$B(4\pi)$\\
\hline
$f_0(1500)$ & $(3.1\pm 0.2\pm 1.1)\times10^{-4}$ & $1.75\times10^{-4}$ &
$1.05\times10^{-4}$&$5.9\times10^{-4}$ \\
\hline
$f_0(1740)$ & $(3.1\pm 0.2\pm 1.1)\times 10^{-4}$ &
$1.73\times 10^{-4}$ & $1.03\times10^{-4}$&$5.9\times10^{-4}$\\
\hline
$f_0(2100)$ & $(5.1\pm0.3\pm1.8)\times10^{-4}$ & $3.08\times10^{-4}$ &
$1.71\times10^{-4}$ & $9.9\times10^{-4}$ \\
\hline
$f_2(1950)$ & ${(5.5\pm 0.3\pm 1.9)\times10^{-4}}$ & $5.58\times10^{-4}$ &
$1.63\times10^{-4}$ & $12.7\times10^{-4}$ \\
\hline
\end{tabular}

\vspace{0.8cm}

Table 2.  Branching ratios of $J/\Psi$ radiative decays to $4\pi$
based on the MARK III data

\vspace{0.3cm}

\begin{tabular}{|l|l|l|l|l|}
\hline
parent meson & $B(\pi^+\pi^-\pi^+\pi^-)$ (measured)
& $B(\pi^+\pi^-\pi^0\pi^0)$ & $B(\pi^0\pi^0\pi^0\pi^0)$&$B(4\pi)$\\
\hline
$f_0(1505)$ & $(2.5\pm 0.4)\times 10^{-4}$ & $1.41\times 10^{-4}$ &
$0.84\times 10^{-4}$&$4.8\times10^{-4}$\\
\hline
$f_0(1750)$ & $(4.3\pm0.6)\times 10^{-4}$ & $2.41\times 10^{-4}$ &
$1.43\times 10^{-4}$&$8.1\times10^{-4}$\\
\hline
$f_0(2104)$ & $(3.0\pm0.8)\times 10^{-4}$ & $1.82\times 10^{-4}$ &
$1.00\times 10^{-4}$&$5.8\times10^{-4}$\\
\hline
\end{tabular}
\end{center}

\vspace{0.8cm}

Using Crystal Barrel results\cite{Scott} and our results here, we get
the relative branching ratio of $Br(f_0(1500)\to 4\pi)/Br(f_0(1500)\to
2\pi)$ to be ($2.1\pm 0.6$), which is compatible with the result
$1.5\pm 0.4$ from $\pi\pi$ scattering phase shifts\cite{D.V.Bugg},
instead of $3.3\pm 0.8$\cite{Scott} by assuming Eq.(1).

In conclusion, our numerical results indicate that interference effects
would make the ratios
$$B(M\rightarrow \pi^+\pi^-\pi^0\pi^0):B(M\rightarrow\pi^+\pi^-\pi^+\pi^-):
B(M\rightarrow\pi^0\pi^0\pi^0\pi^0)$$
much deviating from the naive counting 4:4:1
for isoscalar $0^{++}$ and $2^{++}$ mesons.
The graphs provided in this work can serve as a standard reference
that once one of the $4\pi$ modes
$\pi^+\pi^-\pi^0\pi^0,\pi^+\pi^-\pi^+\pi^-, \pi^0\pi^0\pi^0\pi^0$ is
measured, we can determine the branching ratios of the other modes.

\vspace{0.7cm}
\noindent{\bf Acknowledgment:}
This work is partly supported by the National Natural Science
Foundation of China (NNSFC). We would like to
thank the BES collaboration for supporting their work.
One of us (Zou) thanks D.V.Bugg for useful discussions.

\end{document}